# Analysis of solar gamma rays and solar neutrons detected on March 7th and September 25th of 2011 by Ground Level Neutron Telescopes, SEDA-FIB and FERMI-LAT

Y. Muraki[1)], J. F. Valdés-Galicia[2)], L. X. González[2)], K. Kamiya[3)], Y. Katayose[4)], K. Koga[3)], H. Matsumoto[3)], S. Masuda[1)], Y. Matsubara[1)], Y. Nagai[1)], M. Ohnishi[5)], S. Ozawa[6)], T. Sako[1)], S. Shibata[7)], M. Takita[5)], Y. Tanaka[8)], H. Tsuchiya[9)], K. Watanabe[10)], and J. L. Zhang[11)]

1) ISEE, Nagoya University, Chikusa, Nagoya 464-8601, Japan
2) Instituto de Geofisica, UNAM, 04510, Mexico D.F., Mexico
3) Tsukuba Space center, JAXA, Tsukuba 305-8505, Japan
4) Physics department, Yokohama National University, Yokohama 240-8501, Japan
5) Institute for Cosmic Ray Research, University of Tokyo, Kashiwa 277-8582, Japan
6) Research Institute for Science and Engineering, Waseda University, Tokyo 169-8555, Japan
7) Engineering Science Laboratory, Chubu University, Kasugai 487-0027, Japan
8) Department of Physics, Hiroshima University, Hiroshima 739-8526, Japan
9) Japan Atomic Energy Agency, Tokai, Ibaraki 319-1195, Japan
10) National Defense Academy, Yokosuka 239-8686, Japan
11) Institute of High Energy Physics, Beijing 100049, China

**Abstract**

At the 33rd ICRC, we reported the possible detection of solar gamma rays by a ground level detector and later re-examined this event. On March 7, 2011, the solar neutron telescope (SNT) located at Mt. Sierra Negra, Mexico (4,600 m) observed enhancements of the counting rate from19:49 to 20:02 UT and from 20:50 to 21:01 UT. The statistical significance was 9.7σ and 8.5σ, respectively. This paper discusses the possibility of using this mountain detector to detect solar gamma rays.

In association with this event, the solar neutron detector SEDA-FIB onboard the International Space Station has also detected solar neutrons with a statistical significance of 7.5σ. The FERMI-LAT detector also observed high-energy gamma rays from this flare with a statistical significance of 6.7σ. We thus attempted to make a unified model to explain this data.

In this paper, we report on another candidate for solar gamma rays detected on September 25th, 2011 by the SNT located in Tibet (4,300 m) from 04:37 to 04:47 UT with a statistical significance of 8.0σ (by the Li-Ma method).

## 1. Introduction

The solar flare observed at 19:48 UT on March 7, 2011 may be one of the most important flares observed in Solar Cycle 24. Although the intensity of the flare measured by the GOES satellite was moderate, M3.7, the event has already provided us with much important information. From this information, we were able to point out an emission of long-lasting, high-energy gamma rays detected by the FERMI-LAT satellite. This emission continued for almost 14 hours and high-energy gamma rays with an energy of

~4GeV were recorded.[1] Another relevant fact is that this marked the first detection of solar neutrons in Solar Cycle 24 using the solar neutron telescope (SEDA-FIB) onboard the International Space Station (ISS).[2] Complementary to these observations, a solar neutron telescope (SNT) located at Mt. Sierra Negra in Mexico registered two enhancements of the counting rate during the flare.[3] The Mt. Sierra Negra observatory is located at an altitude of 4,600m a.s.l. In this paper, we will discuss whether these observation results may be explained by a unified model.

This paper is organized as follows: The next section introduces the data observed by the referred instruments. Sections 3 and 4 describe our interpretations that could explain the data consistently. In section 5, we summarize the results.

## 2. Observation Data

Here we present the space environment data obtained from around 19:30 to 21:00 UT on March 7, 2011. Data were obtained using five different satellites and one ground-based detector: GOES, RHESSI, FERMI-GBM,[4] FERMI-LAT,[5] and SEDA-FIB, as well as the SDO and SNT at Mt. Sierra Negra, Mexico. Figure 1 presents the time lines of the different data.

1. *GOES* data

The soft X-ray sensors on the GOES satellite detected a flare that started at 19:43 UT and reached its maximum at 20:12 UT with magnitude M3.7. The flare position was identified at about N23W50 on the solar surface in Active Region 11164.

2. *RHESSI* satellite

The RESSHI satellite also observed this flare from 19:27 to 20:08 UT. The emission of hard X-rays with 25-300 keV started at ~19:47 UT and peaked at 20:00 UT. Two-dimensional plots are made for the times of maxima at 19:57 UT and 20:01 UT using the hard X-ray data of the RHESSI satellite. The peak positions of hard X-rays and soft gamma rays are estimated using the RHESSI data at the solar coordinates of (625”, 560”) for 19:57 UT and (610”, 560”) for 20:01 UT, respectively. Figure 2 presents the plots.

3. *FERMI-GBM* results

The FERMI satellite involves two main detectors (GBM and LAT). The GBM monitors gamma-ray bursts in the range of X-rays between 6 and 300 keV.[6] The FERMI-GBM detector observed the Sun from 20:02 to 20:40 UT on March 7, 2011. The intensity of hard X-rays in the range of 100-300 keV was 2,000 counts per second at 20:02 UT, the data profile shows that the maximum was shortly before observation started; in the second peak at 20:38 UT, a counting rate of ~800 counts per seconds was recorded. Unfortunately, the detectors failed to observe the rising phase of the first impulsive flare. When comparing the time profile of the GBM with that of GOES, however, a clear difference can be recognized. *The observation using soft X-rays only showed only peak bump, in contrast, the observation using hard X-rays clearly showed a second peak at ~20:38 UT*.[6]

4. *FERMI-LAT* results

The high-energy gamma-ray detector FERMI-LAT observed the flare from 20:15 to 20:40 UT. High-energy gamma rays with a peak intensity at 200 MeV were detected. The highest gamma rays reached $E\gamma$~4 GeV. The emission of gamma rays continued for 14 hours, the details of which have been published.[1] The FERMI-LAT detector failed to detect the impulsive phase of the flare.

5. *SEDA-FIB* results on the ISS

The SEDA-AP is a neutron sensor onboard the ISS. The sensor can measure the energy and arrival direction of neutrons. Details may be found in the references.[7-10] The production time may be estimated using information on the energy of neutron-induced protons in the sensor. Figure 3 shows the energy spectra of solar neutrons. The spectra are reduced assuming an instantaneous production of those neutrons. The most probable production time of neutrons is estimated to be around 19:58 UT for the first peak and 20:37 UT for the second peak. The SEDA-FIB also failed to detect the impulsive phase of the first peak. This is because the ISS passed over the night area of the Earth until 20:02 UT.

6. *SDO* satellite

The Solar Dynamics Observatory observes the Sun by means of ultra-violet telescopes over different wavelengths. The SDO satellite successfully observed the emission of a Coronal Mass Ejection (CME) associated with this flare from the start time of emissions. Thus, the satellite not only observed the start of the CME but also its development from the start time of an impulsive flare at ~19:47 UT.[11]

Figure 4 shows several images taken at 19:48 UT, 20:02UT, 20:37 UT, and 20:50 UT by the 30.4-nm SDO/AIA telescope. The 30.4nm line emission is produced by the helium ions. The intensive emission area of hard X-rays determined by the RHESSI observation is circled in the image. 19:47 UT corresponds to the start time of the magnetic loop forming the CME, while 20:50 UT corresponds to the peak time of the counting rate detected by the SNT at Mt. Sierra Negra. Thus, 19:58 UT and 20:37 UT are the estimated solar neutron production times. Figure 5 shows the SDO/AIA images at 19:47 UT, 19:58 UT, and 20:38 UT at other wavelengths**.**

7. *Solar Neutron Telescope* at Mt. Sierra Negra

Figure 6 shows the time profile of the L1 channel between 19:00 and 21:30 UT as observed by the SNT at Sierra Negra (SN-SNT). The SNT consists of plastic scintillators surrounded by proportional counters (PC) in anticoincidence, to separate the flux of neutral particles from charged particles. The L1 channel measures the flux of neutral particles (n or gamma) crossing the scintillator and triggering the PC underneath [12].

Two clear enhancements from 19:49 to 20:02UT and from 20:50 to 21:01UT may be recognized. Figure 7 shows the time profile of the neutron channel, N1, that is the flux of neutral particles measured at the scintillator. The N1/L1 was 0.16~0.40, depending on the counting rate of the N1 channel. The ratio sharply contrasts from the solar neutron event observed on September 7, 2005 using the same detector [13]. The ratio was ~7 for the event of September 7, 2005.

8. *Solar Neutron Telescope* at Tibet

We found a similar event in the data obtained by the SNT [14] located in Tibet (4,300 m a.s.l.) on September 25th, 2011, following the M7.4 flare at 04:33 UT. Figure 8 shows the time profile of L3 channel between 04:00 and 07:00UT. In comparison with the neutron channel, N1, $E_n$ >40MeV (Figure 9) , a clear enhancement is recognized. The statistical significance is 8.0 $\sigma$ by the Li-Ma method. Figure 10 presents the time profile of the neutron monitor located in Tibet at the same time.

Under the scintillator of the SNT in Tibet, between the Layer 1 and 3, a 10cm thick wood plate is installed for the selection of high energy neutrons. The wood plate absorbs low energy protons induced by low energy neutrons. The density of the wood plate is about 1.0 gr/cm$^3$. This facility was not added to the SNT at Mt. Sierra Negra. According to the Monte Carlo calculation based on the GEANT 3 program [15], the detection efficiencies of solar neutrons and solar gamma rays of N1, L1 and L3 channels are 0.2, 0.07, 0.0001 for neutrons, and 0.3, 0.1, 0.01 for gamma rays respectively. These numbers are given for incident neutral particles with an energy of 150MeV. Thus the L1 and L3 channels have an excellent ability to discriminate between neutrons and gamma rays. We also plot the time profile of the counting rate of one minute value of the L1 channel of Tibet SNT at the same time in Figure 11. Figure 12 plots the L3/N1 ratio. Except around 4.63UT (or 04:38UT), L3 and N1 data variated similarly.

Figure 13 shows the energy spectrum of solar neutrons assuming instantaneous production at 04:45:00 UT. It seems that has two components, forming the low energy bump and the high energy tail. The assumption that these neutrons were produced at the same time maybe does not correctly reflect the case of this flare.

In Figure 14, we present images of the solar surface around the maximum time of the flare taken by the SDO/AIA UV telescope. The image indicates two bright spots. Therefore, the gross feature of the flare observed by the X-rays may be explained by the superposition of the two independent flares that developed with slight different time lag. The long lasting flare beyond 20 minutes may reflect this circumstance and the energy spectrum of Figure 12 may be described by this assumption.

## 3. Comparison of Observed Intensities between SEDA and SNT

At the time of the flare on March 7, 2011, the SNT at Mt. Sierra Negra (SN-SNT) and the SEDA-FIB recorded two enhancements.[15] This raises the question of whether both enhancements were produced in association with the M3.7 flare. The enhancements recorded by the SNT were observed by the L1 channel that consists of proportional counters located just under the scintillators. Except for a minor, non-significant, enhancement of the anti-counter, a positive signal of neutrons was not recorded. A similar trend can be recognized in the data observed at Tibet for the 25 September 2011 event. Therefore, these enhancements recorded by the SNTs must have been produced by gamma rays. In such case, we have estimated the intensity of gamma rays that entered the top of the atmosphere.

The counting rate of the L1 channel at Mt. Sierra Negra was estimated as 152 counts/m$^2$/min. and 145 counts/m$^2$/min. for the first and second peaks, respectively. Monte Carlo (MC) calculations were used to estimate the incident flux at the top of the atmosphere. According to the MC calculation, the attenuation is estimated as 0.1 for photons with energy of E = 1 GeV, 0.01 for photons of 200MeV, 0.003 for 100MeV, and 0.001 for photons with energy of 30 MeV. Here we take the attenuation factor as 0.001. Then the flux at the top of the atmosphere is estimated to be 150,000 counts/m$^2$/min.

The flux of solar neutrons with $E_n$ > 30 MeV is conversely estimated as follows: From Fig. 3 (left), the number of neutrons after “decay correction” is estimated as 149

counts/100 cm$^2$ or 14,900/m$^2$. Given the SEDA-FIB sensor's efficiency in detecting neutrons (estimated at 2%), the net value at the top of the atmosphere after decay correction turns out to be $7.5 \times 10^5$ counts/m$^2$. If these neutrons were emitted during seven minutes, the intensive gamma ray emission time, we could compare both values. Then we would find that both are of the same order at a ratio of about 0.71 (= neutrons/gamma rays). We have checked these values obtained by the recent MC calculation based on GEANT 4 [16].

Using the MC calculation to estimate the flux in the direction toward the Earth, we assumed an opening angle of 57 degrees between the vertical direction and the direction toward the Earth. The MC calculation predicts the ratio to be 0.7 at $E_{n\ or\ \gamma} > 30$ MeV in case the accelerated protons have the power index of -4.0. Although the SEDA did not observe the impulsive phase, it could measure neutrons produced during the impulsive phase, as the arrival time of neutrons near the Earth is expected to be 7-20 minutes later than that of the photons. By comparing the energy spectrum of the first peak at 19:58 UT (Fig. 3, left) with the spectrum of the second peak at 20:37UT (Figure 3, right), the latter seems to have a harder spectrum.

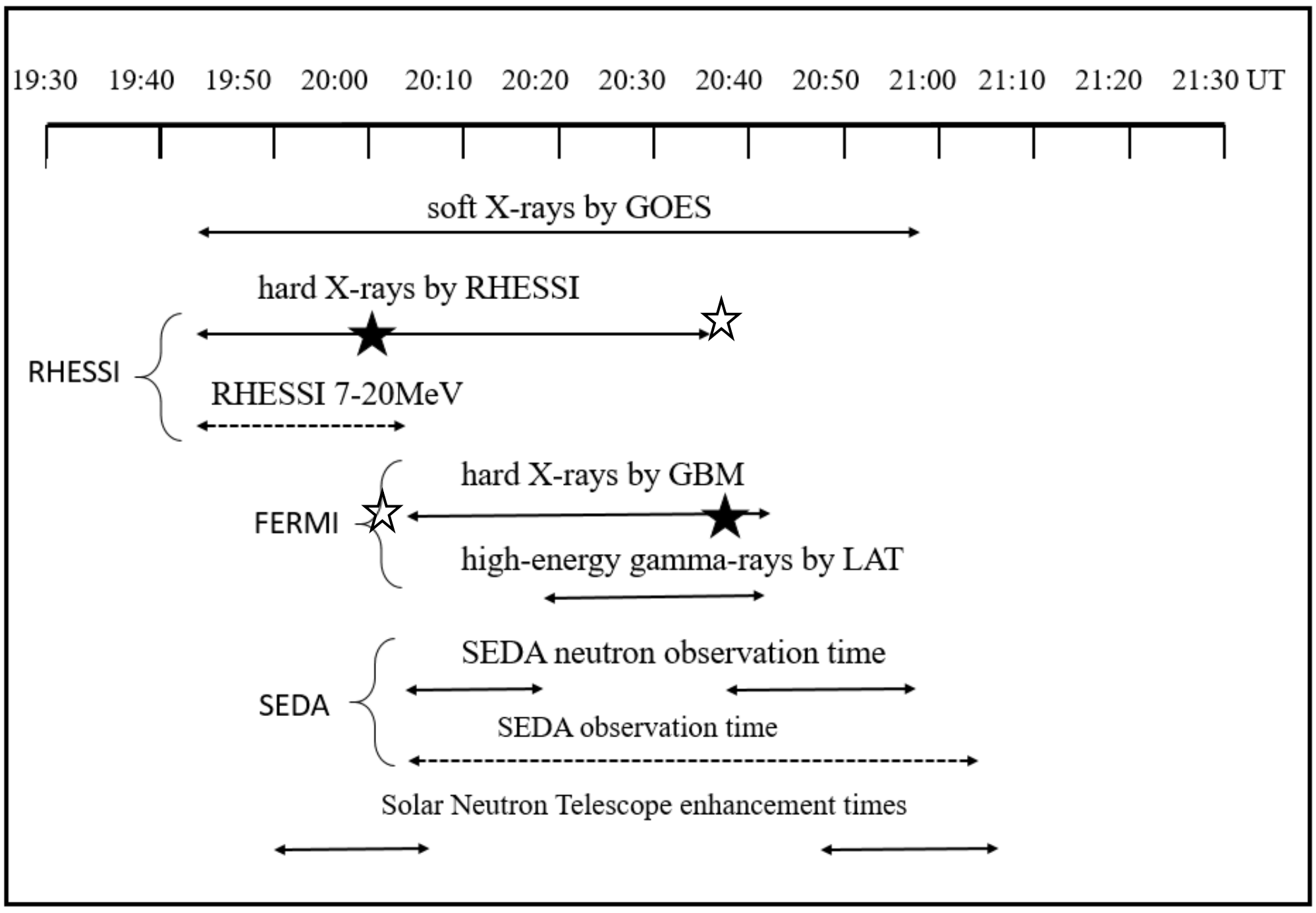

Figure 1. The event time profile of each detector as a function of universal time during 19:30–21:30 UT on March 7, 2011. The full star mark ★ denotes the peak times of hard X-rays. The open star (☆) represents the peak that respective detector failed to see.

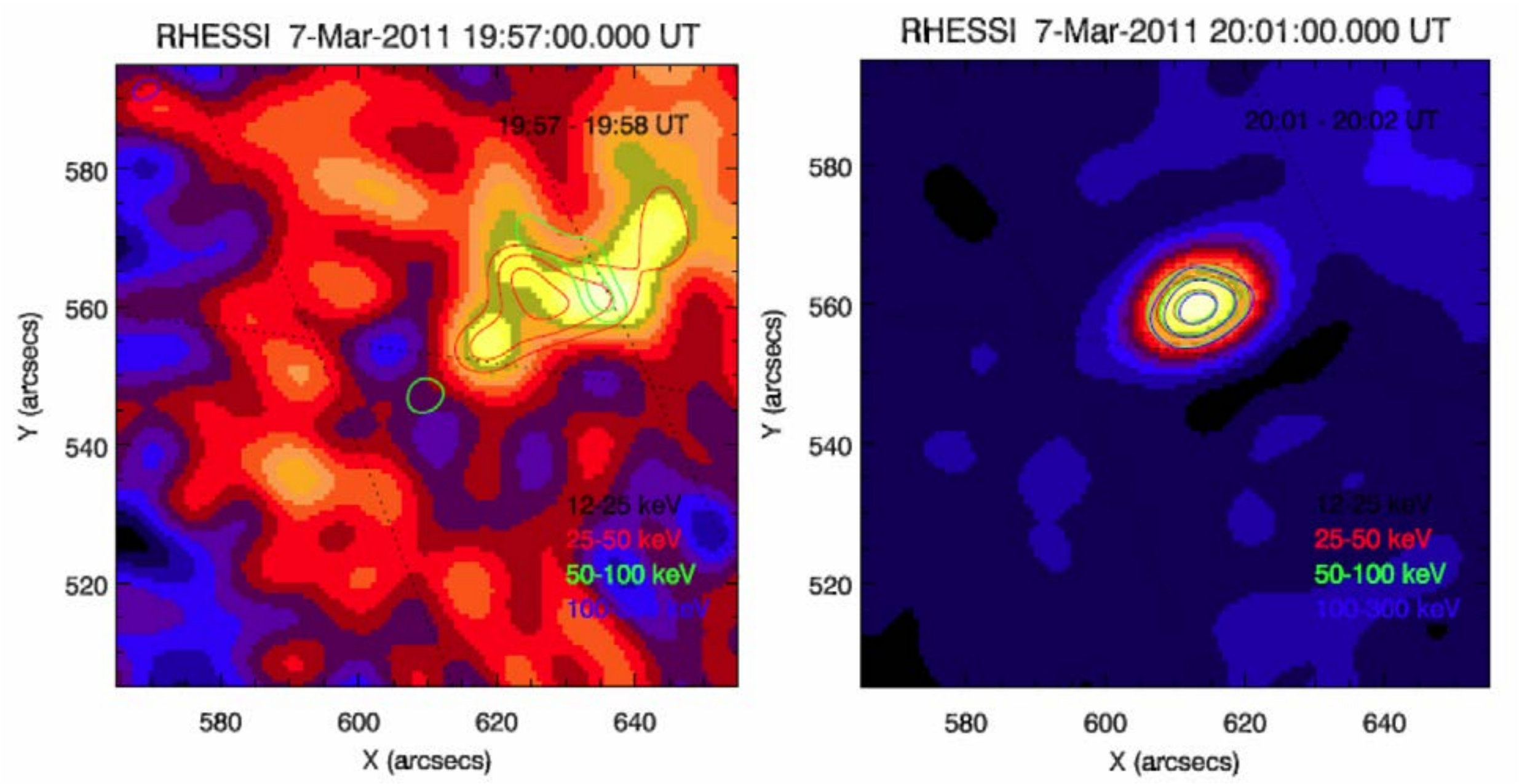

Figure 2. Density maps of X-rays observed by the RHESSI satellite at (a) 19:57-19:58 UT and (b) 20:01–20:02 UT. The red contours represent the intensity of hard X-rays with 20–50 keV; the green contours correspond to X-rays with energy of 50–100 keV.

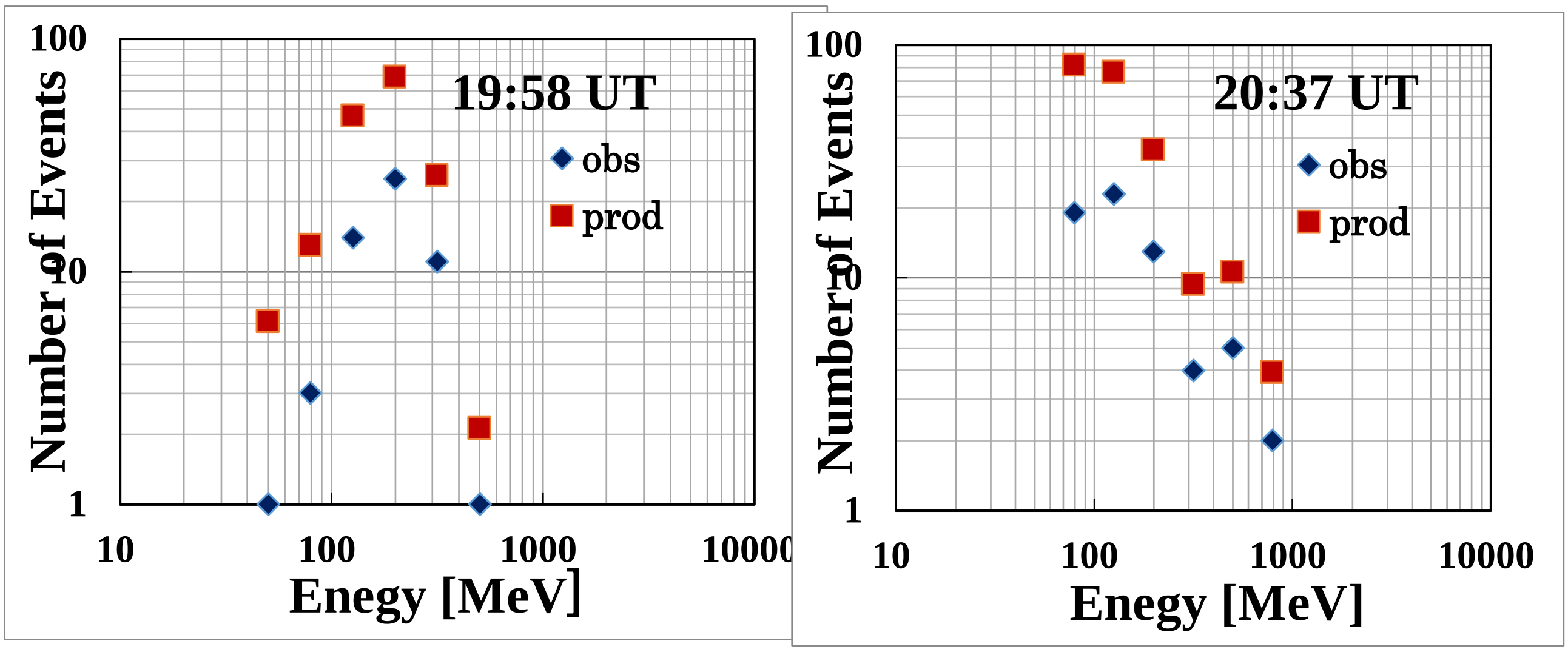

Figure 3. Energy spectra of solar neutrons for the 1st bump and 2nd bump on SEDA=NEM detector. Here we assume that neutrons were produced instantaneously at 19:58 UT and 20:37 UT when soft gamma rays of 100-300 keV showed maximum intensity. The blue diamond shows the raw data, while the red square represents the corrected data of the decay effect in flight between the Sun and the Earth.

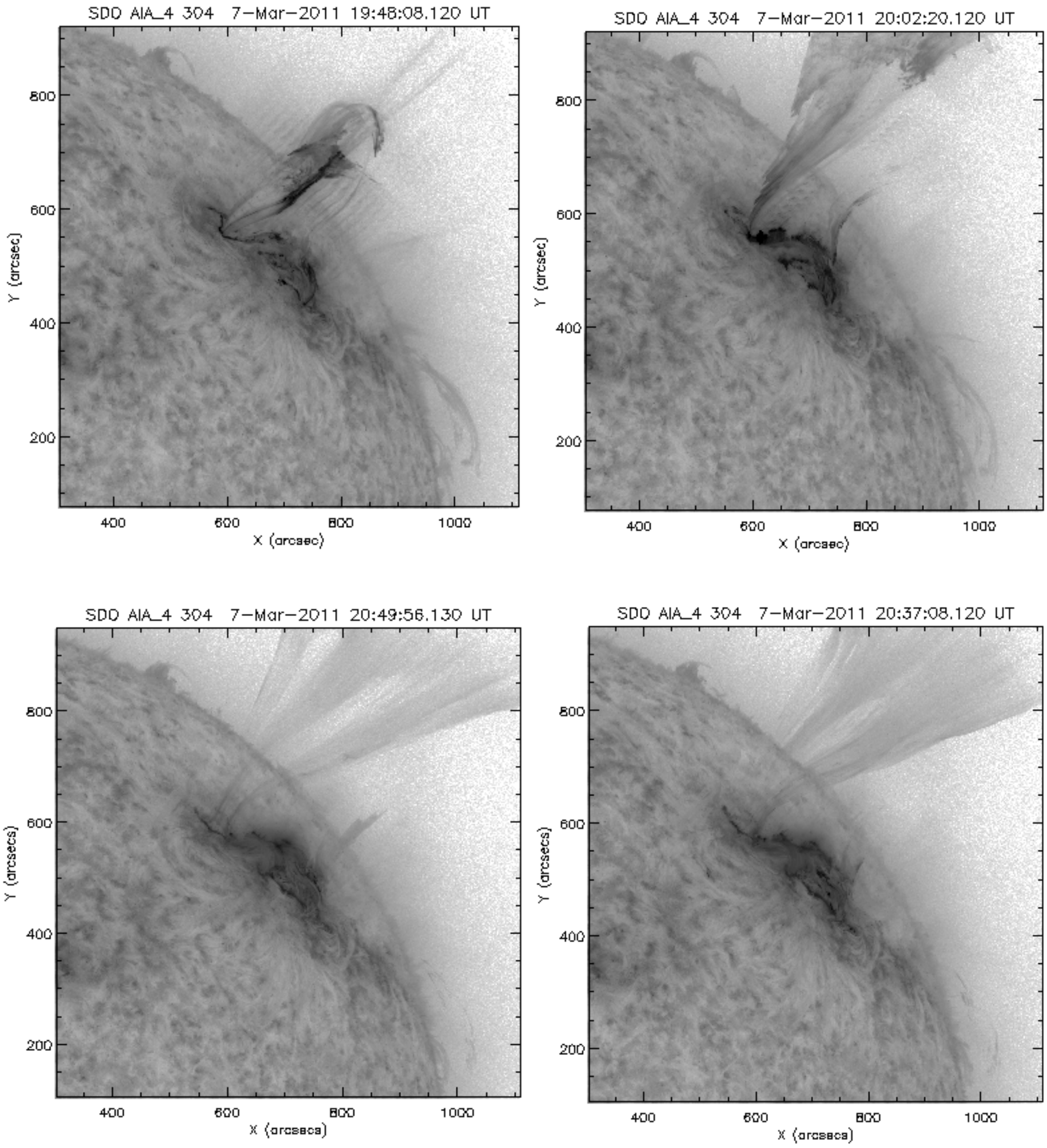

Figure 4. Images of the solar surface around AR1164 taken by the SDO/AIA telescopes at a wavelength of 30.4 nm (clockwise) at 19:48 UT, 20:02 UT, 20:37 UT, and 20:50 UT. The circle indicates the region where the most intensive hard X-rays were observed by the RHESSI satellite.

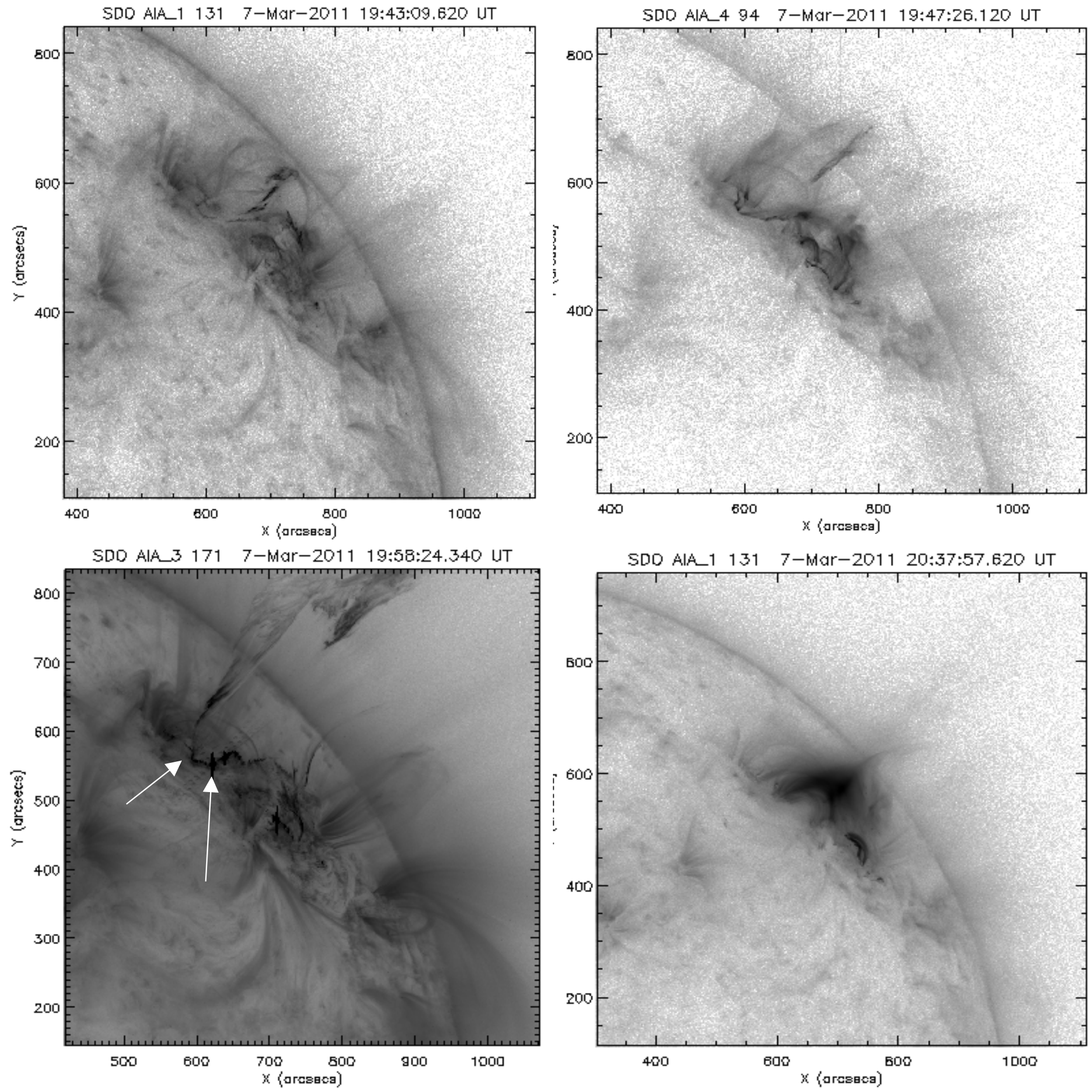

Figure 5. Images of the solar surface taken by the SDO/AIA telescopes with a wavelength of 13.1 nm at 19:43 UT (top-left), 9.4 nm at 19:47:26 UT (top-right), 17.1 nm at 19:58:24 UT (bottom-left), and 13.1 nm at 20:50 UT (bottom-right). At 19:58:24 UT, a flash can be recognized at 625” and 555” (the right arrow). Those hot particles may be transferred into the foot point of the CME at 595” and 560” (indicated by the left arrow), and then become seed particles for high-energy accelerated protons possibly located inside the shock of the CME (not in the figure). Chen et al. estimated the start time of the CME at 19:43 UT [11].

## 4. Comparison of Observed Intensities between SNT and FERMI-LAT

Let us compare the counting rate observed by the FERMI-LAT detector with the counting rate observed by the Sierra Negra SNT. The gamma-ray flux near the Earth measured by the FERMI-LAT around 20:15 UT is estimated as ~57 counts/m$^2$/min. for $E_\gamma$ > 100 MeV.[1] In contrast, the gamma-ray flux of $E_\gamma$ > 30 MeV estimated from the counting rate measured using the SNT is 150,000 counts/m$^2$/min. The flux of gamma rays estimated from SNT observation is about 2,500 times higher than that estimated from FERMI-LAT observation.

One reason arises from the difference of the energy threshold ($E_\gamma$ > 100 MeV and > 30MeV). Another reason for this difference may be due to the fact that the FERMI-LAT detector did not observe the impulsive peak. In the flare on June 11, 1991, Rank et al.[17] and Kanbach et al.[18] reported that the intensity of gamma rays at the impulsive phase was about ~1,000 times more intensive than that of the gradual phase. Therefore, the SNT located on a high mountain may have observed gamma rays emitted during the impulsive phase. However, the SNT could not observe an enhancement of gamma rays emitted during the gradual phase due to the high background.

## Summary and Conclusion

We have reevaluated the time profiles of high energy photons and neutrons observed by several detectors in association with the M3.7 flare on March 11, 2011. The different intensities recorded by the SN-SNT and SEDA may be consistently explained as gamma rays by a new MC calculation.

The excess observed on September 25th, 2011 by the Tibet SNT may be also explained by the detection of solar gamma rays. To the best of our knowledge, there was no report on the detection of solar gamma rays by a ground based detector. This report may be *the first report* on the detection of solar gamma rays. Many cosmic ray detectors are located at high mountains. Gamma rays are a useful tool to study particle acceleration processes at the Sun.

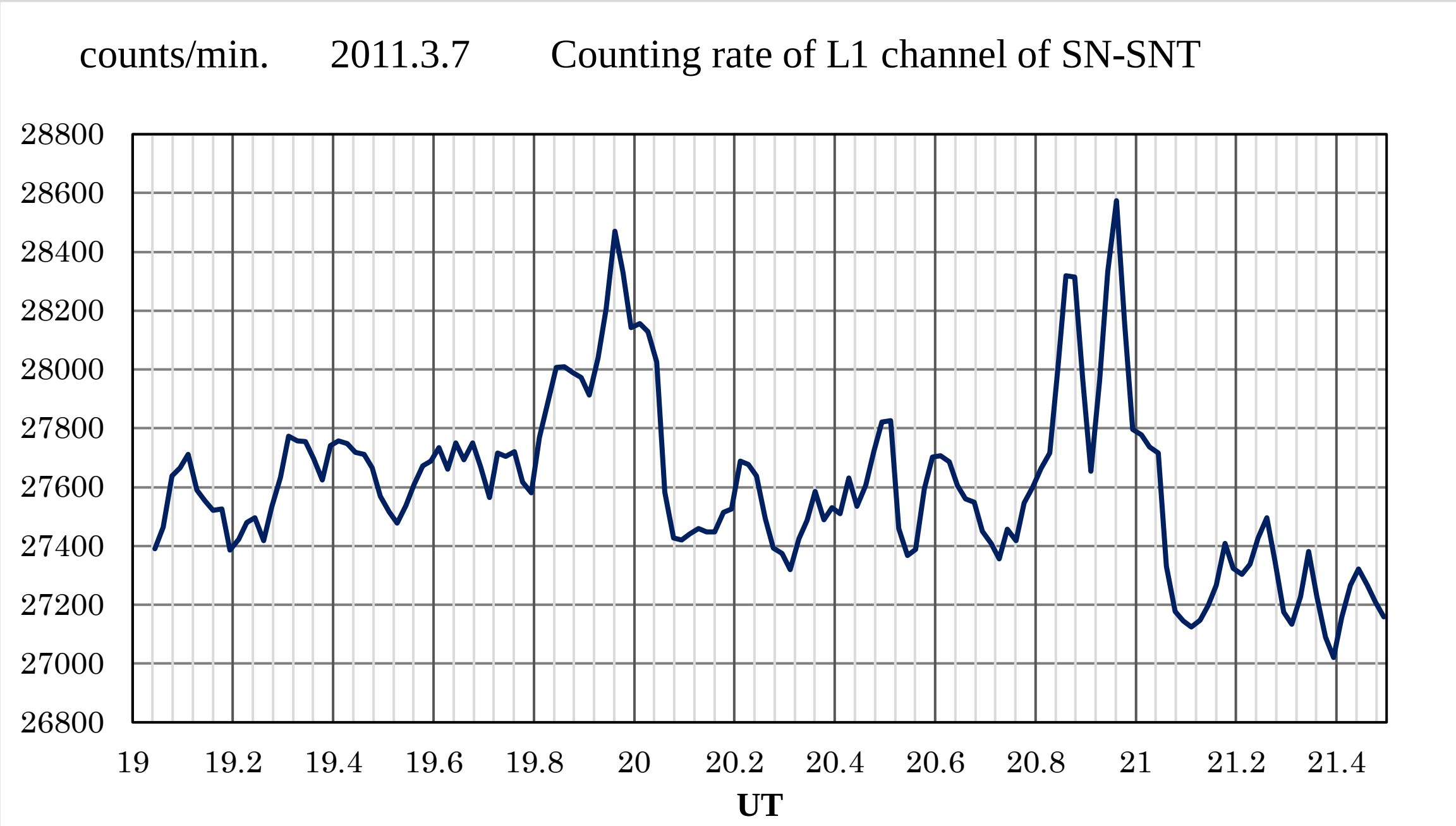

Figure 6. Three-minute running average of one-minute values of the L1 channel observed by the SN-SNT. The counting rate from 19:05 to 21:30 UT is shown. Two enhancements of the counting rate are recognized at 19h49m−20h02mUT (or 19.82-20.03UT) and 20h50m−21h01mUT. The statistical significance is very high at 9.7σ for the 1st peak and 8.5σ for the 2nd peak. As an average background rate, we use 27,500 counts per minute.

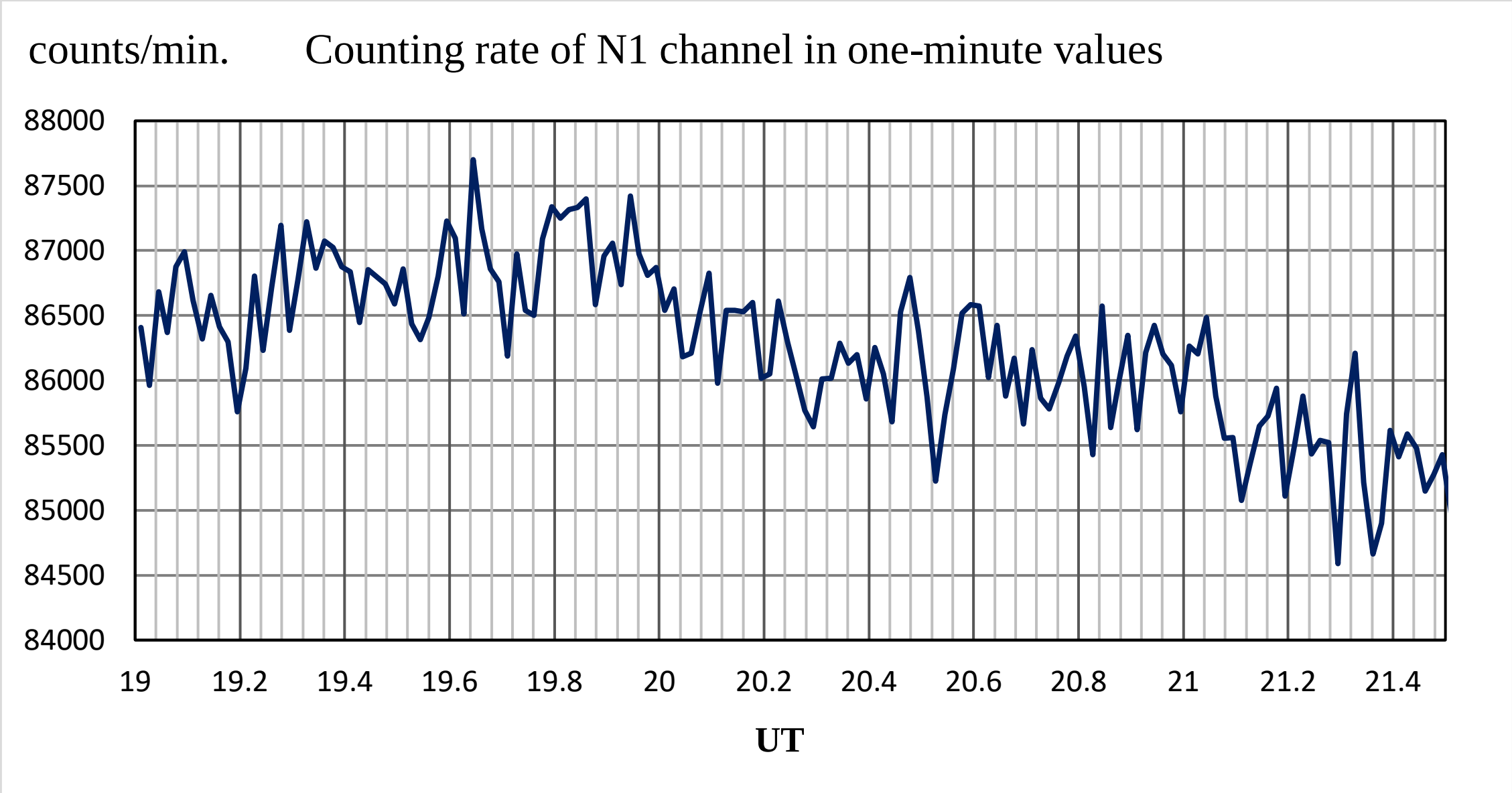

Figure 7. The counting rate of neutron channel N1 of SN-SNT. The vertical counting rate corresponds to the one-minute value for the 4-m$^2$ area scintillator. The N1 channel records neutral incident particles with energy higher than 30 MeV. The energy of the detector has been calibrated by using the energy deposited by passing muons. As seen in the above figure, no remarkable enhancements are recognized around 19.95 UT and 20.9 UT.

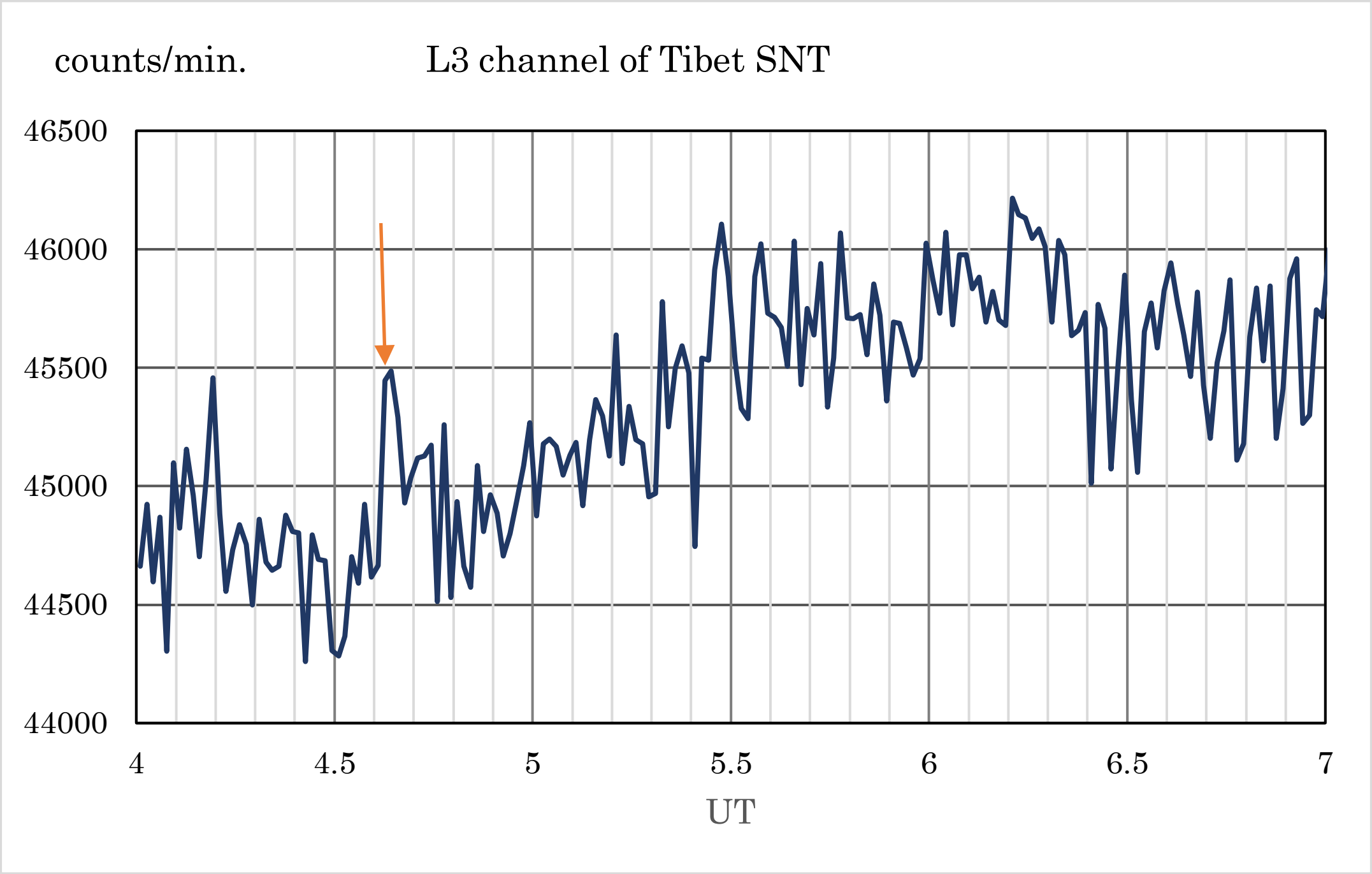

Figure 8. Similar enhancement observed by the SNT located in Tibet (4,300 m). The SNT has an area of 9 $m^2$ and the thickness of the plastic scintillator is 40-cm. The enhancement started at 04:37 UT (12:37 Beijing time (indicated by an arrow). During this time, no enhancement was observed by the neutron monitor located at the same site. Therefore, we think that the enhancement was produced by gamma rays.

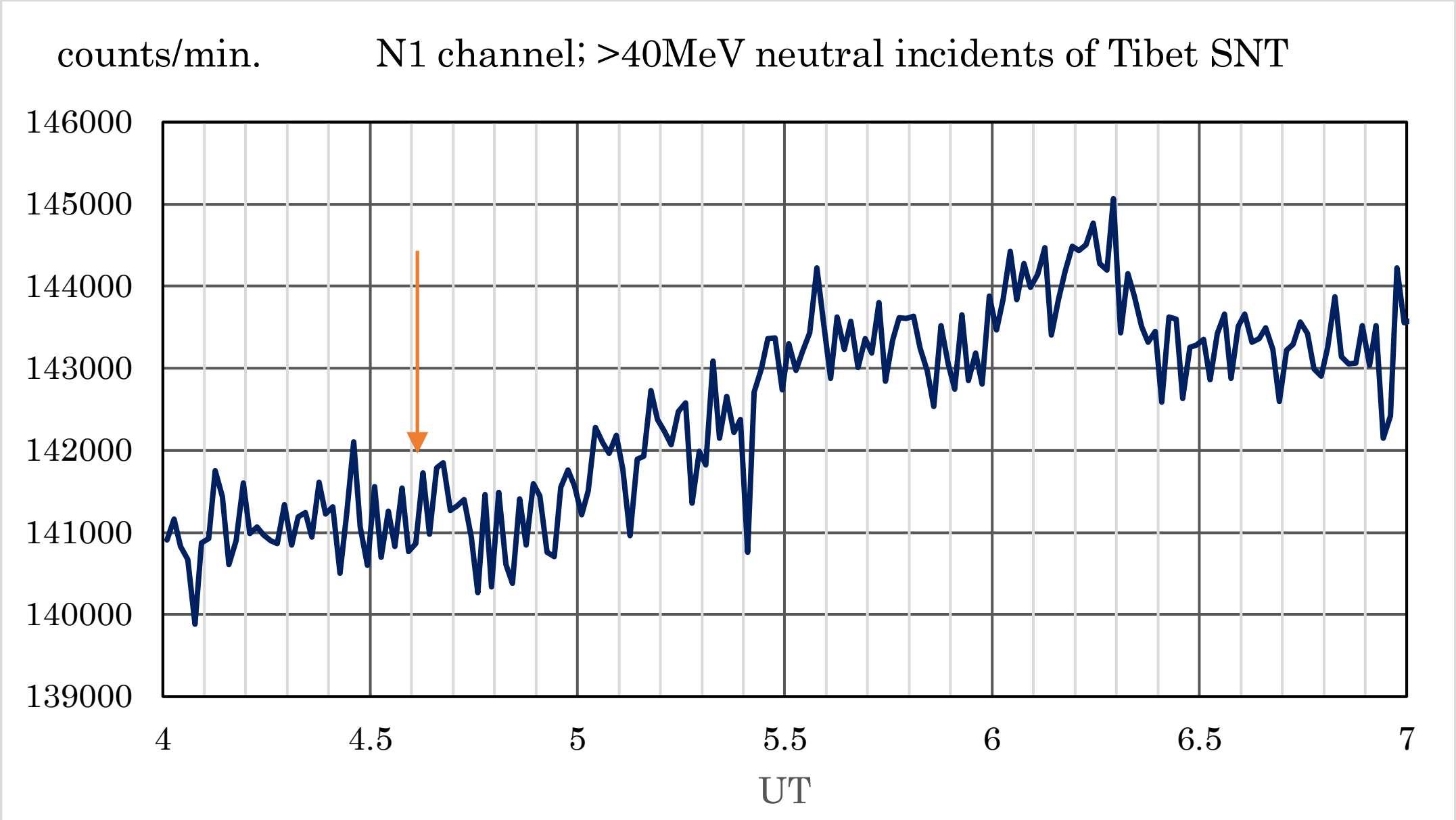

Figure 9. Counting rate of the neutral channel with the threshold energy higher than 40 MeV. Those signals are mainly produced by the protons induced by incoming neutrons via charge exchange process. However when gamma-rays enter into the scintillator, they also produce electron-positron pairs. Electrons and positrons are the minimum ionizing particles, so they penetrate the scintillator and enter into the PR-counters located underneath. Therefore the enhancement of Figure 8 is induced by gamma rays.

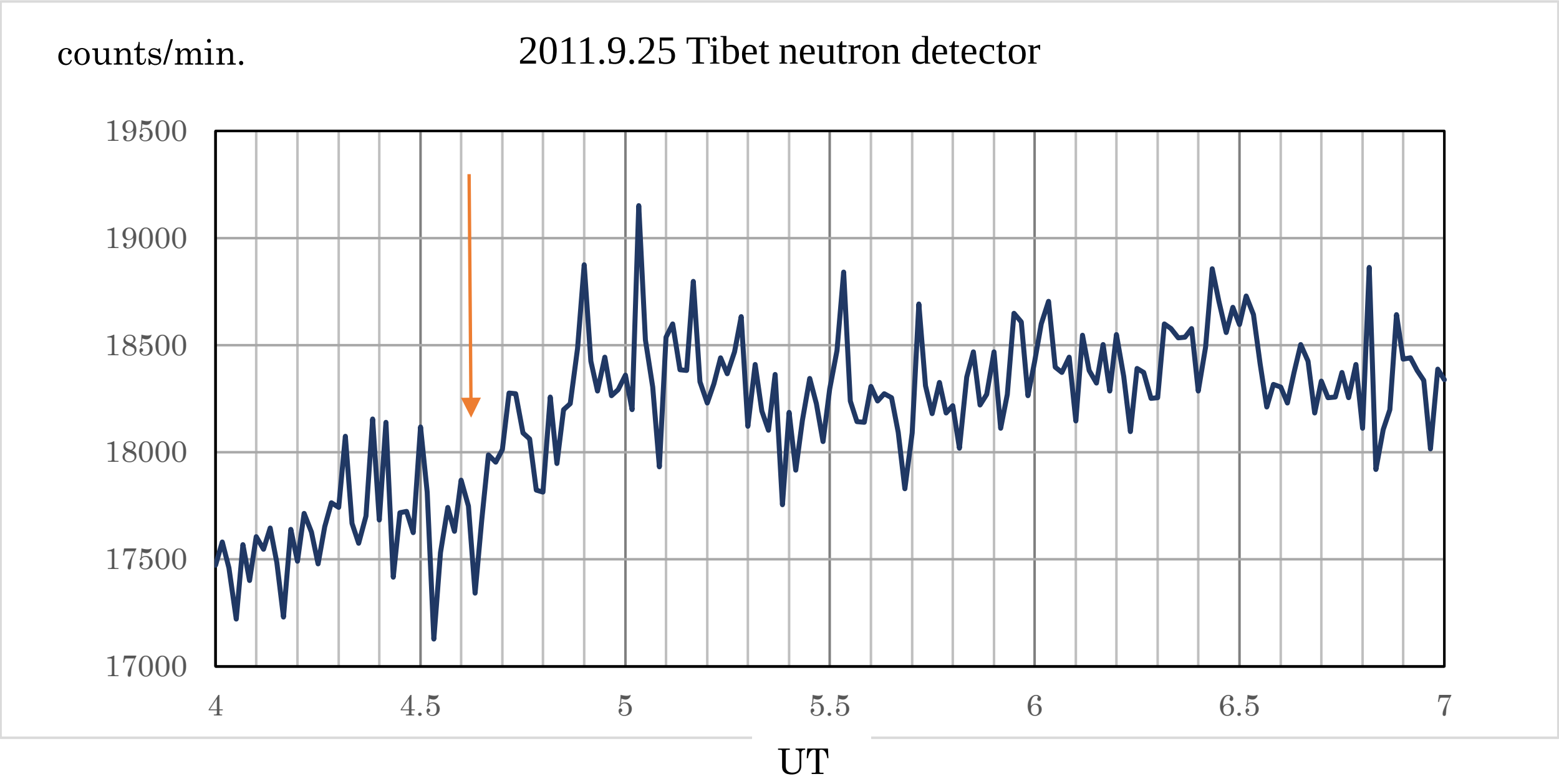

Figure 10. The neutron-muon monitor located at the same site in Tibet (4300 m) [19]. No remarkable enhancement can be seen around 04:37 UT (indicated by the arrow). The data are corrected for the pressure effect.

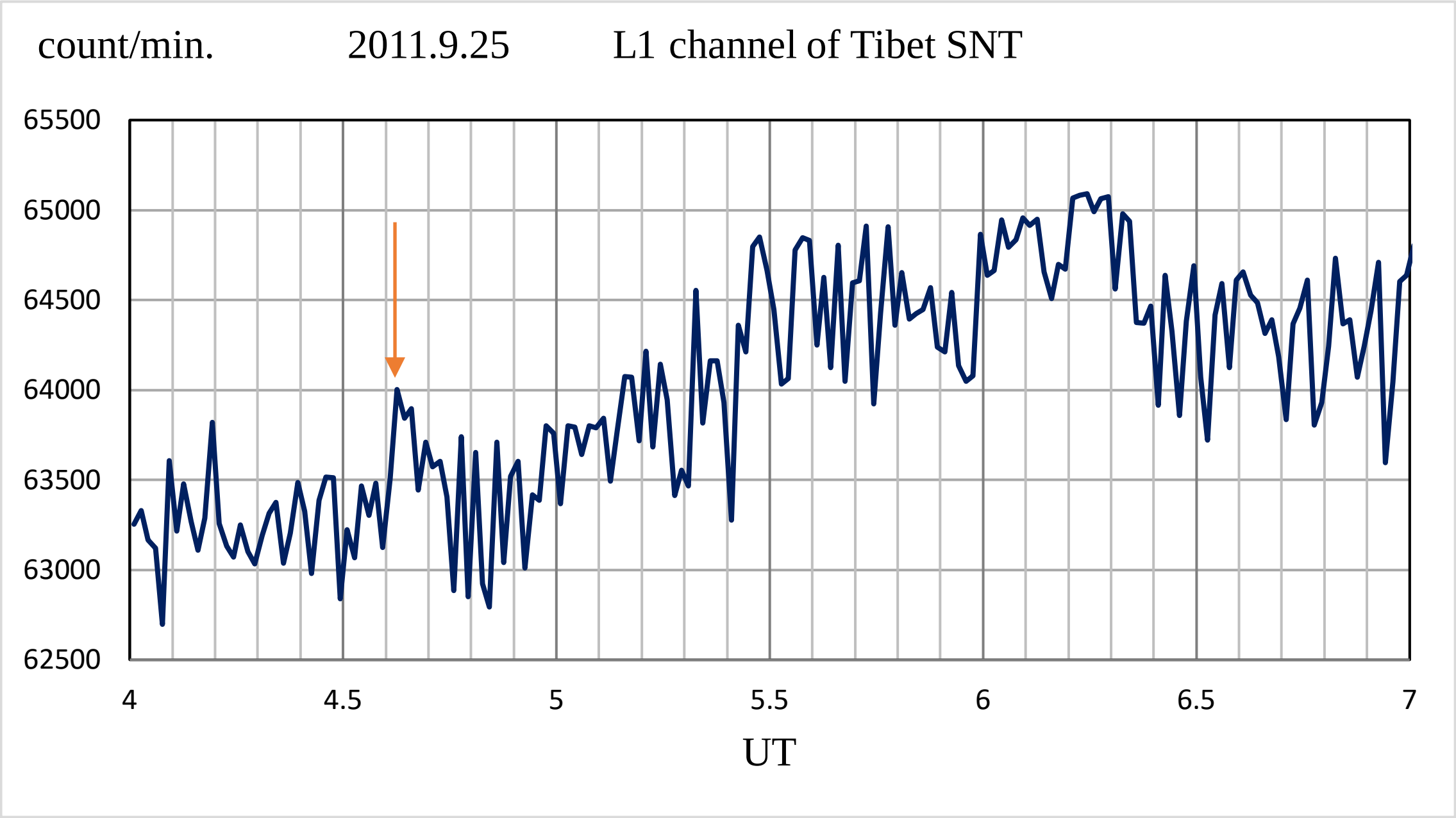

Figure 11. The one minute value of the L1 channel of Tibet SNT. An enhancement may be recognized at 04:37 UT (or 4.62 UT). The counting rate of the L1 channel represents the penetrating particles from the scintillator after neutral particles are converted into charged protons or electrons in the scintillator. Between the L1 channel and L3 channel, a 10cm thick wood plate is installed. The L1 channel has a sensitivity for lower energy protons or electrons in comparison with that of L3 channel. When we take the average counting rate of L1 channel as 63,300, the statistical significance of the enhancement during 04:37 and 04:43UT turns out as to be 8.9σ by the Li-Ma method.

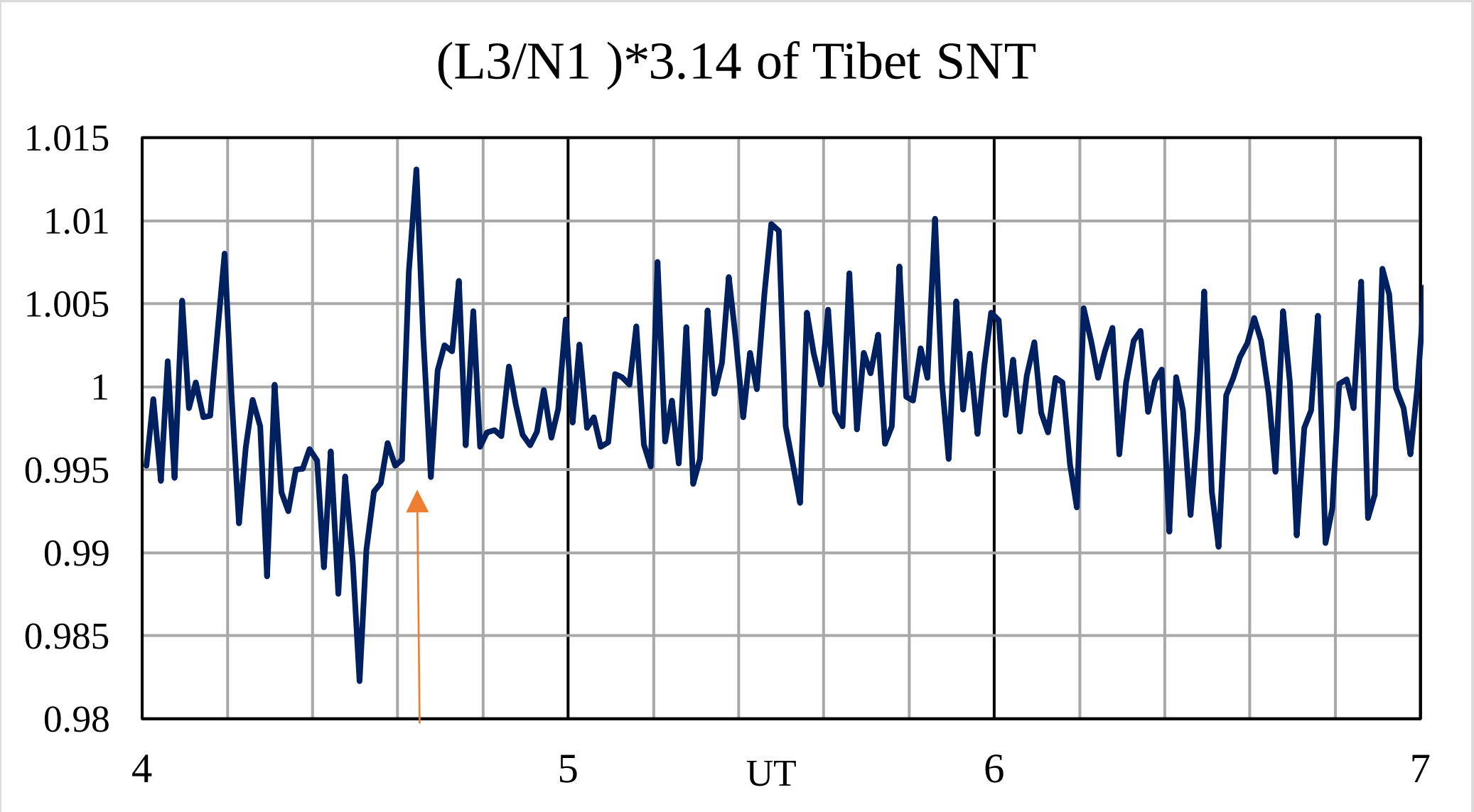

Figure 12. The counting ratio between L3 channel and N1 ($E_n$>40MeV) channel. To the counting rate of L3, a factor 3.14 is multiplied. Except around 4.63UT (or 04:38UT), the L3 and N1 channels varied with the statistical fluctuation.

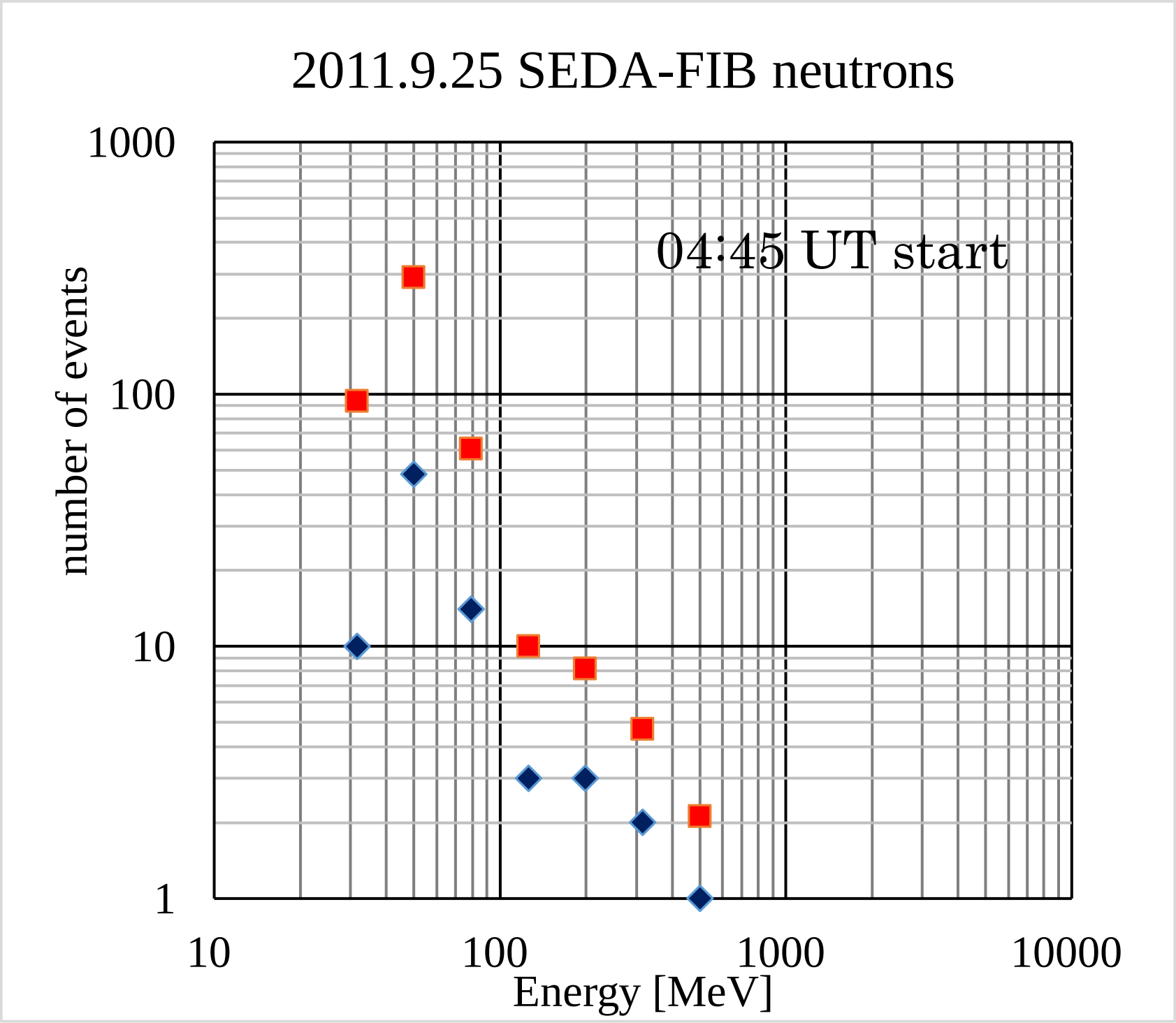

Figure 13. The energy spectrum of solar neutrons by the SEDA-FIB detector. The spectrum is obtained assuming an instantaneous production at 04:45UT. The blue diamond shows the raw data, while the red square represents the corrected data of the decay effect in flight between the Sun and the Earth. It seems that the spectrum may be composed by the two components, the low energy part and high energy part. This means that the assumption of one instantaneous production time may be imperfect. The peaks of the neutron monitor around 5.1 UT may reflect the arrival of solar neutrons (See Figure 10).

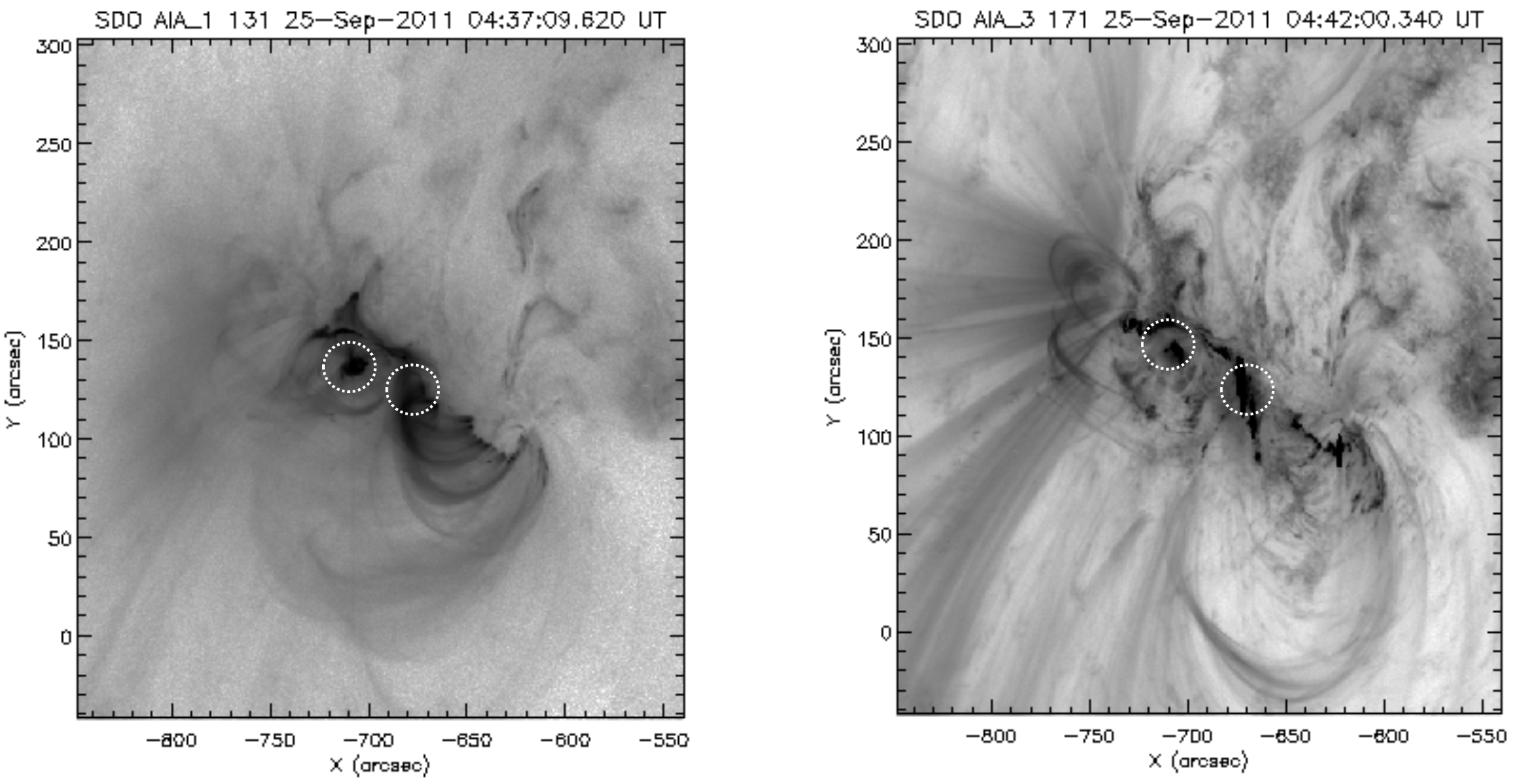

Figure 14. The SDO/AIA images on September 25, 2011 event. The dotted white circles present the locations of the maximum intensity of the UV light. Left: 04:32:09UT, right 04:42-00UT. The coordinates are (-670”, 130”) and (-705”, 140”) respectively.

## References

[1] Ackermann, A. et al. (2014) ApJ 785, 15, doi:10.1088/0004-637X/787/1/15.
[2] Koga, K. et al. (2017) Solar Physics (to be published).
[3] Muraki, Y. et al. (2013) submitted to Proceedings of the 33rd ICRC in Rio, arXiv:1307.5125v1 [astro-ph] (19 Jul 2013).
[4] Meegan, C. et al. (2009) ApJ 702, 791, GBM, doi:10.1088/0004-637X/702/1/791.
[5] Atwood et al. (2009) ApJ, 697, 1071, DOI:10.1088/0004-637X/697/2/1071.
[6] http://sprg.ssl.berkeley.edu/~tohban/browser/?show=grth+qlpcr
[7] Imaida, I. et al. (1999) Nucl. Inst. Meth., A421, 99, doi:10.1016/S0168-9002(98)01163-2.
[8] Koga, K. et al. (2011) Astrophy. Space Sci. Trans., 7, 411, DOI: 10.5194/astra-7-411-2011.
[9] Muraki, Y. et al. (2012) Advances in Astronomy. Article ID 37904, DOI: 10.1155/2012/379304.
[10] Muraki, Y. et al. (2016) Solar Physics. 291, 1241, DOI: 10.1007/s11207-016-0887-0.
[11] Chen, X. et al. (2013) ApJ, 763, 43, doi:10.1088/0004-637X/763/1/43.
[12] Valdes-Galicia, J.F. et al. (2004) Nucl. Inst. Meth. A535, 656, DOI:1 0.1016/j.nima.2004.06-148.
[13] Sako, T. et al., ApJ, 651 (2006) L69.
[14] Muraki, Y. (2007) Astroparticle Physics. 28, 119-131, DOI: 10.1016/j.astrophysics.2007.04.012.
[15] Watanabe, K., Proceeding of the Cosmic-Ray Research Section of Nagoya University, Vol. 46 (2005) 1-247. ISSN 0910-0717 (PhD thesis in English).
[16] Kamiya, K. et al, Proceeding of 35th ICRC (Pusan).
[17] Rank, G. et al. (2001) Astronomy and Astrophysics. 378, 1046, DOI: 10.1051/0004-6361:20011060.
[18] Kanbach, G. et al. (1993) Astronomy & Astrophysics, Suppl. 97, 349.
[19] Zhang, J.L. et al., Proceeding of 29th ICRC (Pune), 1(2005), 9-12. On-line version is available: http://articles.adsabs.harvard.edu/full/seri/ICRC./0029//A000009.000.html
Or: https: //cds.cern.ch/record/957287/files/11009-chn-zhang-J-abs2-sh11-oral.pdf